\documentclass[prl,showpacs,preprintnumbers,amsmath,amssymb]{revtex4}

\usepackage{graphicx}
\usepackage{dcolumn}
\usepackage{bm}

\begin{document}

\title{Non-linear interaction in random matrix models of RNA}

\author{Itty Garg}
\author{Pradeep Bhadola}
\author{N. Deo}
 \email{ndeo@physics.du.ac.in}
\affiliation{Department of Physics and Astrophysics, University
of Delhi, Delhi 110007, India}
\date{\today}

\begin{abstract}
A non-linear Penner type interaction is introduced and studied
in the random matrix model of homo-RNA. The asymptotics in length of the partition function
is discussed for small and large $N$ (size of matrix). The interaction
doubles the coupling ($v$) between the bases and the dependence of the
combinatoric factor on ($v,N$) is found. For small $N$, the effect of interaction changes
the power law exponents for the secondary and tertiary structures. The specific heat
shows different analytical behavior in the two regions of
$N$, with a peculiar double peak in its second derivative for $N=1$ at low
temperature. Tapping the model indicates the presence of multiple solutions.
\end{abstract}

\pacs{02.10.Yn, 02.70.Rr, 05.40.-a, 87.10.-e}

\maketitle

In the fundamental understanding of RNA folding combinatorics, the
study of exact enumeration of RNA secondary structures with
crossings (pseudoknots) is an important ongoing research direction
\cite{1}. In this context, Orland and Zee
proposed a random matrix-field theoretic model \cite{2} which
addressed the problem of exact RNA structure combinatorics which
with certain simplifying assumptions \cite{3} enumerated all
possible planar and non-planar structures. In this model, the idea
of introducing external linear interaction was explored in
\cite{4,5,6} with the objective to observe structural and other
statistical changes (if any). Even with this simple interaction, it
was shown that the interaction imposed additional constraints on the
RNA chain which could physically imply changes in temperature,
applied pressure, proximity with ions \cite{6}. So, within this
matrix model framework, an important interaction to study is the
Penner type that may capture interesting properties such as: effects
of interactions with more complex molecules and biomolecules,
multiple solutions, frozen (glassy) states \cite{7}. The Penner
matrix models appear in the context of disordered systems \cite{8}
(initially to calculate correct fluctuations for the conductance),
string theory \cite{9} (to get accurate critical exponents for
quantum gravity) and spin glasses \cite{10} (as mappings to high temperature
$p$-spin glasses). In this letter studying the Penner type
interactions increases the moduli space of structures
from ${\cal M}_{g,0}$ to ${\cal M}_{g,n}$ where $g$ is the genus
of the surface and $n$ gives the number of faces or punctures of the Riemann
surfaces. Thus the generalized partition function of RNA matrix model (of length $L$:
$(i,j)=1,....,L$) with a logarithmic interaction in the action is given by

\begin{equation}
Z_{L}(N) = \frac{1}{A_{L}(N)} \int \prod_{i=1}^L d\phi_{i}
e^{-\frac{N}{2} \sum_{i, j = 1}^L (V^{-1})_{i, j} Tr \phi_{i}
\phi_{j}} e^{\frac{N}{2} \sum_{i = 1}^L (W^{-1})_{i} Tr
[\log \phi_{i}^{2}]} \frac{1}{N} Tr \prod_{i = 1}^L (1 + \phi_{i}),
\end{equation}

where $\phi_{i}$'s are $L$ independent ($N \times N$) random
(symmetric) hermitian matrices placed at each base site in the chain
with the interactions contained in $V_{ij}$ (see \cite{2,3,4} for
all other notations and conventions). The normalization constant is given by

\begin{equation}
A_{L}(N)=\int \prod_{i=1}^{L}
d\phi_{i} e^{-\frac{N}{2}\sum_{i, j=1}^L (V^{-1})_{i,j} Tr \phi_{i}
\phi_{j}} e^{\frac{N}{2} \sum_{i=1}^L (W^{-1})_{i} Tr [\log
\phi_{i}^2]}, 
\end{equation}

and $\frac{1}{N} Tr \prod_{i} (1+\phi_{i})$ is the characteristic observable of the
model. With the simplifications, $V_{ij}=v$ and $W_{i}=w$, and a
series of Hubbard Stratonovich Transformations \cite{4}, Eq. (1)
becomes 

\begin{equation}
Z_{L}(N)=\frac{1}{A^\prime_{L}(N)} \int d\sigma e^{-N Tr
[\frac{1}{2v}\sigma^{2} - \frac{1}{2w} (\log \sigma^{2})]}
\frac{1}{N} Tr (1 + \sigma)^L,
\end{equation}

where $\sigma$ is an ($N \times N$) matrix, the potential is given by
$V(\sigma)=[\frac{\mu}{2}\sigma^{2}-\frac{t^{\prime\prime}}{2}\log
\sigma^{2}]$ (with $\mu=1/v$, $t^{\prime\prime}=1/w$) and the normalization is

\begin{equation} 
A^\prime_{L}(N)=\int d\sigma e^{-N Tr
[\frac{1}{2v}\sigma^{2}-\frac{1}{2w}\log \sigma^{2}]}.
\end{equation}

These potentials are of the Gaussian Penner matrix models \cite{9} and
will be solved along those lines. The spectral density $\rho_{N}(z)$
of the matrix $\sigma$ at finite $N$ is

\begin{equation}
\rho_{N}(z)=\frac{1}{A^\prime_{L}(N)} \int d\sigma e^{-N Tr
V(\sigma)} \frac{1}{N} Tr \delta(z-\sigma), 
\end{equation}

where $z$ are the
eigenvalues of $\sigma$. Defining $G(t,
N)=\sum_{L=0}^{\infty}Z_{L}(N)\frac{t^L}{L!}$ as the exponential
generating function of the partition function \cite{3,4} and using
the identity $\int_{-\infty}^{+\infty}dz\rho_{N}(z)=1$ gives

\begin{equation}
G(t,N)=\int_{-\infty}^{+\infty} dz \rho_{N}(z) exp^{t (1 + z)}. 
\end{equation}

To solve $G(t,N)$, the expression for spectral density is found using
the orthogonal polynomial method (Deo, \cite{9}). For these models,
the orthogonal polynomials are given by ${\cal P}_{n}(z)=z^{n}+l.o.$
which satisfy the orthogonality condition,

\begin{equation}
\int_{-\infty}^{+\infty}dz e^{-N V(z)} {\cal P}_{n}(z){\cal
P}_{m}(z)=h_{n}\delta_{nm}. 
\end{equation}

For the symmetric Gaussian Penner
matrix model, orthogonal polynomials split into even and odd
polynomials. The even set obeys the orthogonality condition

\begin{equation}
\int_{0}^{\infty}dy e^{-N^\prime[\nu_{0}(y)-t^\prime\log y]}{\cal
P}_{n}(y){\cal P}_{m}(y)=h_{n} \delta_{nm}, 
\end{equation}

where $y=z^{2}$, $N^\prime=N/2$, $\nu_{0}(y)=2V_{0}(z)=\mu y+....$ and
$t^\prime=(t^{\prime\prime}-\frac{1}{2N^\prime})$. The odd ones obey

\begin{equation}
\int_{0}^{\infty}dy e^{-N^\prime[\nu_{0} y-{\bar t}^\prime\log y]}
{\bar {\cal P}}_{n}(y) {\bar {\cal P}}_{m}(y)={\bar h}_{n}
\delta_{nm},
\end{equation}

where ${\bar t}^\prime=(t^{\prime\prime}+\frac{1}{2N^\prime})$. It is sufficient
to work with either of the two polynomials as each one can
completely determine the recursion coefficients independent of the
other. The normalized even orthogonal polynomials from solving the
orthogonality relations are 

\begin{equation}
\psi_{2n}(y)=e^{-\frac{N^\prime}{2}
[\mu y-t^\prime \log y]} \hat{{\cal P}}_{n}(y)
=\left[\frac{n!(N^\prime \mu)^{N^\prime
t^\prime+1}}{\Gamma(n+1+N^\prime t^\prime)}\right]^{1/2}
y^{\frac{N^\prime t^\prime}{2}} e^{-\frac{N^\prime \mu y}{2}}
L^{N^\prime t^\prime}_{n}(N^\prime \mu y), 
\end{equation}

which in the case of odd polynomials has ${\bar t}^\prime$ instead of $t^\prime$. The kernel for this function is
defined by $K(y_{i},y_{j})=\sum_{n=0}^{N-1} \psi_{2n}(y_{i})
\psi_{2n}(y_{j})$ which gives the normalized spectral density for
the even polynomials as $\rho^{e}_{N}(y)=\frac{1}{N} K(y,y)$ (the
superscript $e$ represents even). Thus the spectral density for
large $N$ limit is 

\begin{equation}
\rho^{e}_{N}(z^{2})=\frac{1}{N} \sum_{n=0}^{N-1}
\left[\frac{n!(N^\prime \mu)^{N^\prime
t^\prime+1}}{\Gamma(n+1+N^\prime t^\prime)}\right] z^{2N^\prime
t^\prime} e^{-N^\prime \mu z^{2}} [L^{N^\prime
t^\prime}_{n}(N^\prime \mu z^{2})]^2, 
\end{equation}

where $y=z^{2}$ is considered. Using the relation $\rho^{e}_{N}(z)=z \rho^{e}_{N}(z^{2})$ (Tan; Deo
in \cite{9}) and substituting $\rho^{e}_{N}(z^{2})$ in the
generating function for the even polynomials, Eq. (6) gives

\begin{table}
\caption{The Table lists even partition functions for $L$
upto 6 for the matrix model with logarithmic interaction.}
\begin{tabular}{lll}
\hline \hline
$L$ & $Z^{e}_{L}(N)$ \\
\hline
1 & 1 \\
2 & $1+v(2+\frac{1}{N})$ \\
3 & $1+3v(2+\frac{1}{N})$ \\
4 & $1+6v(2+\frac{1}{N})+2v^{2}(4+\frac{3}{N})+\frac{v^{2}}{N^{2}}$ \\
5 & $1+10v(2+\frac{1}{N})+10v^{2}(4+\frac{3}{N})+\frac{5v^{2}}{N^{2}}$ \\
6 & $1+15v(2+\frac{1}{N})+30v^{2}(4+\frac{3}{N})+5v^{3}(8+\frac{8}{N}+\frac{1}{N^{3}})$ \\
      & $+(15v^{2}+20v^{3})/N^{2}$ \\
\hline \hline
\end{tabular}
\end{table}

\begin{equation}
G^{e}(t,N) = \frac{1}{N} \sum_{n=0}^{N-1}
\left[\frac{n!(N^\prime \mu)^{N^\prime
t^\prime+1}}{\Gamma(n+1+N^\prime t^\prime)}\right]
e^{\frac{t^{2}}{4N^\prime \mu}+t} 
\int_{-\infty}^{+\infty} \frac{dx}{\sqrt{N^\prime \mu}}
\left[\frac{x}{\sqrt{N^\prime \mu}}\right]^{2N^\prime t^\prime+1}
e^{-\left[x-\frac{t}{2\sqrt{N^\prime
\mu}}\right]^{2}} \left[L^{N^\prime t^\prime}_{n}(x^{2})\right]^{2}.
\end{equation}

To solve this integral, a simplification is made, $N^\prime
t^\prime=1/2$ ($-1/2$ for the odd) which requires
$t^{\prime\prime}=2/N$ ($-2/N$ for the odd). Substituting this and
using the relation $H_{2n+1}(x)=(-1)^{n} 2^{2n+1} n! x
L^{1/2}_{n}(x^{2})$ (for odd polynomials, $H_{2n}(x)=(-1)^{n} 2^{2n}
n! L^{-1/2}_{n}(x^{2})$) in Eq. (12) gives 

\begin{equation}
G^{e}(t,N)=\frac{1}{N}
\sum_{n=0}^{N-1} \left[\frac{e^{\frac{t^{2}}{4N^\prime
\mu}+t}}{\Gamma(n+\frac{3}{2})}\right] \frac{1}{n! (2^{2n+1})^{2}}
\int_{-\infty}^{+\infty} dx e^{-\left[x-\frac{t}{2\sqrt{N^\prime
\mu}}\right]^{2}}\\\times\left[H_{2n+1}(x)\right]^{2}. 
\end{equation}

Using the formula $[H_{k}(x)]^{2}=\sum_{l=0}^{k} \frac{(k!)^{2}
2^{k-l}}{(l!)^{2} (k-l)!} H_{2l}(x)$ and the integral
$\int_{-\infty}^{+\infty} dx e^{-(x-y)^{2}} H_{n}(x)=\sqrt{\Pi}
y^{n} 2^{n}$, $G^{e}(t,N)$ becomes

\begin{equation}
G^{e}(t,N) = \frac{1}{N} e^{\frac{t^{2}}{4N^\prime \mu}+t}
\sum_{n=0}^{N-1} \sum_{l=0}^{2n+1} \binom{2n+1}{l}
\frac{t^{2l}}{l! (2N^\prime\mu)^{l}},
\end{equation}

which differs from odd polynomials in $\sum_{l=0}^{2n}
\binom{2n}{l}$. The partition functions can be obtained from Eq. (13)
(Table I). The general form of the partition function for any
$N^\prime t^\prime$ requires a rigorous mathematical analysis and is
left for a future work. In this model, $N$ plays a dual role of
contributing to the strength of the external interaction and a genus
identification parameter of the structures. However with $N^\prime
t^\prime=(1/2,-1/2)$, the genus characterization cannot be extracted
systematically. In the model of \cite{2,3}, the structures (and
their genus characterization) were obtained on the moduli space of a zero puncture
Riemann surface ${\cal M}_{g,0}$. Constructing the matrix model of
RNA with a logarithmic interaction generalizes the study of RNA
structures to that of $n$-punctured Riemann surfaces,
${\cal M}_{g,n}$. The Euler characteristics for these models are given
by $(V-E+n)=(2-2g)$ where $V$ and $E$ give the number of vertices
and edges respectively and includes the additional factor of faces or
punctures $n$. The genus characterization of the structures
obtained from the model is therefore changed from \cite{3}.
For this model the Feynman diagrams are given by the
fat-graphs \cite{9}.

\section{Asymptotic Analysis of the partition function} 

The asymptotic behavior of the partition function at large length is found
numerically as in \cite{5}. The analysis is divided into two parts,
(I) the estimation of structure combinatoric factor and (II) the
determination of (secondary and tertiary) power law
exponents with $L$.

\subsection{Combinatorics}
The combinatorics of the structures is given by
$X=(2\sqrt{f(v,N)}+1)^{L}$ where (i) $X=3^{L}$ for the model in
\cite{3} and (ii) $X=(3-\alpha)^{L}$ in \cite{5} (here $f(v,N)=v=1$
in both the cases). The analysis is done for lengths upto
$L=(40,80,160)$. For each length, different $v$ $(1$ to $6)$ are
considered and for each $v$ different values of $N$ (1 to
100000) are chosen so as to observe the effect of interaction. In
order to determine the form of $f(v,N)$ for this model,
$ln[Z^{e}_{L}(N)]$ is plotted with $L$ for different $(L,v,N)$. The linearly fitted
slopes and intercepts are found. For a given
$v$,\\ 
(i) $f(v,N)$'s for each $N$ are found from the slopes using the
expression $f(v,N)=(\frac{e^{Slope}-1}{2})^2$ and \\
(ii) $f(v,N)$'s hence found are plotted linearly with $1/N$. 

Therefore for each $v$, a functional form of $f(v,N)$ is obtained in terms of $1/N$. Then
for each length, the slopes and intercepts from the functional forms
of $f(v,N)$ for different $v$ are plotted as a function of $v$ to
obtain the final $f_L(v,N)$ expression. For the log interaction the
form is found to be $f^{log}(v,N)=\left[2v_{oz}(1+\frac{a_{oz}}{2N}) \right]$
where $a_{oz}\sim16$ for $L=160$ and for the $oz$ model it is
$f^{oz}(v,N)=\left[v_{oz}(1+\frac{a_{oz}}{N}) \right]$. This results
in two observations: (i). The base pairing interaction strength
$v_{oz}$ in \cite{3} is doubled. This implies, $v$ in the log
interaction model is re-scaled to twice the $v$ in the $oz$ model
which is similar to the way in which the parameter $\alpha$ in the
linear interaction model \cite{6} re-scaled $v$ to
$v/(1-\alpha)^{2}$. (ii). An additional dependence of the
combinatoric factor on $N$ is found ($va/N$) for all small and large
values of $N$ for the log interaction and $oz$ models (which comes out to be same). Such a
dependence has not been found before in these matrix models.

\begin{figure}
\includegraphics[width=6cm]{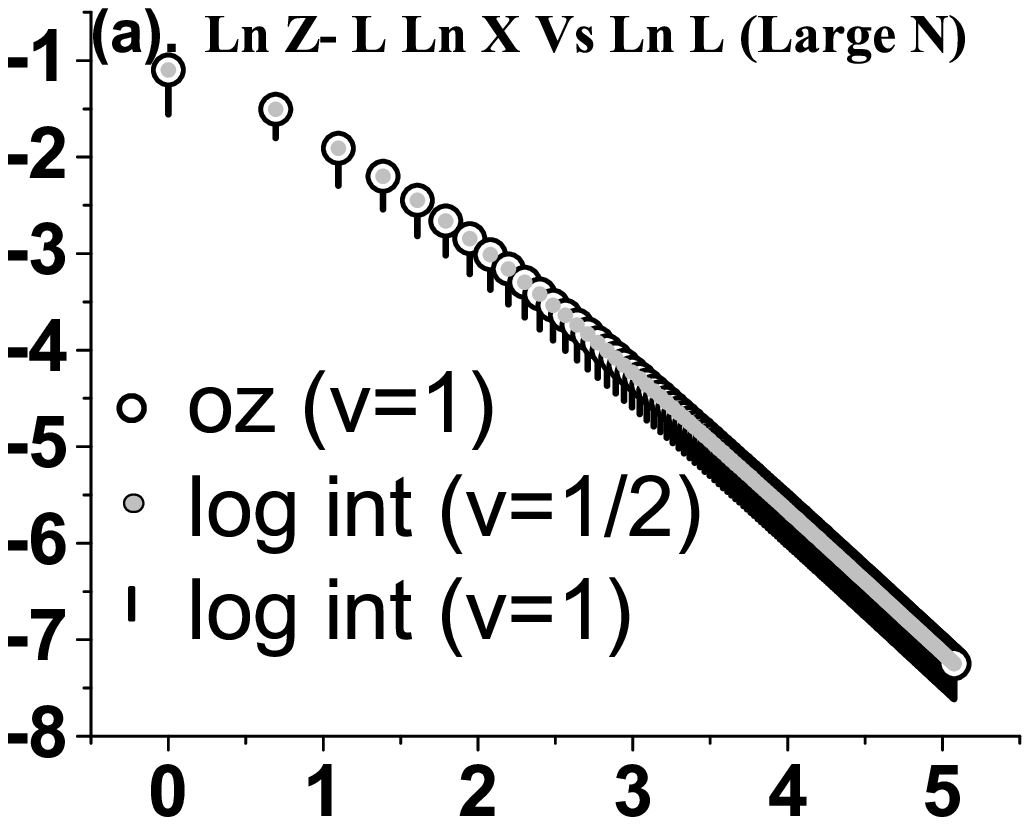}
\includegraphics[width=6cm]{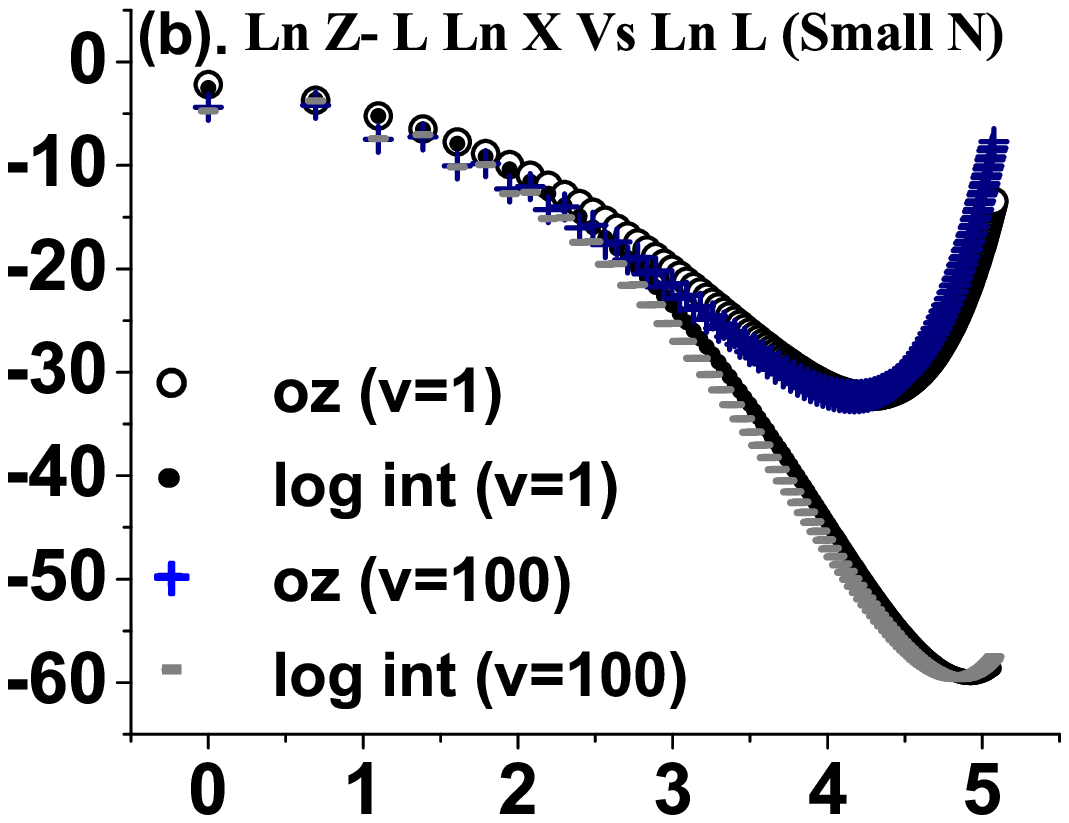}
\caption{$Ln Z^{e}_{L}(N)-L ln X$ is plotted with $ln L$ for the log
interaction and $oz$ models for (a). $N=1000000$ and (b). $N=1$.}
\end{figure}

\subsection{Power law exponents}
In order to extract the exponents corresponding to secondary
and tertiary contributions, the $ln L$ dependence in the partition
function is studied. This analysis is done for $L=160$ and is
divided into small $N$ ($N=1$) and large $N$ ($N=100000$) regions.

\subsubsection{Large $N$}
To study the secondary exponent, ($ln Z-L ln
X$) is plotted with $ln L$ (Fig1(a)). (i) The plot is a straight
line with slope $-1.37$ which approaches $-1.5$ when larger and
larger values of $N$ are considered. This implies at large $N$, the
major contribution is from the genus zero (secondary) structures and
the power law $L^{-3/2}$ holds true as for the $oz$ and linear
interaction models \cite{3,4,5,6}. (ii) For $v=1$, the $oz$ and log
interaction curves are distinctly different. On substituting $v=1/2$
in the log partition function, the two model curves coincide. So for
large $N$ and $v=1/2$, the structures in the log interaction model
reduce exactly to the structures found in the $oz$ model. Thus the
effect of interaction at large $N$ is visible in terms of $v$ only.
(iii) The slopes of the curves for different $v$'s are nearly
$-1.37$ implying that the slope (i.e. the exponent $-3/2$) is
independent of $v$. (b). Next, to study the tertiary exponent, ($ln
Z-L ln X+(3/2) ln L$) is plotted with $L$ which shows: (i) the
structure of the curves for the two models is exactly the same. So
the genus contribution in the large $N$ limit has the same form for
both the models. (ii) The extrapolation of curves on y axis gives
the value of $k_{0}$ (for large $N$, see \cite{3}). For $v=1$, these
are different for the two models showing that the coefficients
$k_{g}$'s will be different. For large values of $v$, the points for
the even and odd lengths (at small lengths) split up for both the
models (also seen at small $N$). The splitting is more for log model
and may be due to different $k_{g}$'s which seem to depend upon $v$.

\subsubsection{Small $N$}
($Ln Z-L ln X$) is plotted with $ln L$ (Fig 1(b))
and the following observations are made: (i) The effect of
interaction in the small $N$ region is due to $v$ and $N$ where $N$
contributes dominantly while the contribution of $v$ is very little.
(ii) The different $v$ curves for the two models clearly indicate
that no value of $v$ will ever reduce the log model to $oz$ (or vice
versa). This is a vital difference between the two models to
establish their uniqueness particularly at small $N$. (iii) The plot
is no longer a straight line with slope $-1.5$ but a U shaped curve
which also includes the contribution from crossed structures. (iv)
The length at which secondary contribution becomes less dominant
than tertiary (given by minima of the curve) is larger for the log
interaction model. Therefore the effect of interaction is mainly on
the secondary structures for a given length \footnote{The
interpretation that first half portion of the U shaped curve
corresponds to zero genus structures and the latter half to higher
genus ones, comes from $\sum_{g}L^{3g-3/2}$ \cite{3}.}. The large number of structures
obtained from the asymptotic analysis of the partition function
can be attributed to the increased structure space of all the $n$-punctured
Riemann surfaces ${\cal M}_{g,n}$.

\section{Specific Heat}
The specific heat is defined as $C_{(v=L)}=-T(\frac{\partial^{2}F}{\partial T^{2}})$ where $F$ is
the total free energy of the polymer chain for a given length $L$.
The calculations (in Fig. 2) are performed for the log interaction
and $oz$ models for $L=160$ and $N=(1,3,6,10,100000)$ with
$v=e^{-\epsilon/kT}$ (where Boltzmann constant and base specific
binding energy ratio $\epsilon/k=1$). The following observations are
made: (i) $C_{v}$ with $T$: For a fixed length, the peak value is
maximum for $N=1$ and decreases (almost half) as larger values of
$N$ are considered for the two models. The peak $C_{v}$ decreases
upto certain value of $N$ after which it becomes nearly constant
however large is the value of $N$ for both the models. (ii)
$dC_{v}/dT$ with $T$: The first derivative of specific heat shows a
kink for small $N$ (for both the models) whereas for large $N$'s, no
such behavior is seen. Further, the kink shifts to smaller
temperatures as length is increased. (iii) $d^{2}C_{v}/dT^{2}$ with
$T$: At small $T$'s, the curve shows (unusual) double peaks for
$N=1$ for the log interaction model. There is a systematic
conversion of the double peak into becoming a single peak as $N$ is
increased slowly. For large $N$, there is a peculiar kink present in
the lower part of the curve for both the models. The kink is
slightly more pronounced in the log interaction model than the $oz$.
A similar such kink is visible (for largest length considered
$L=1024$) in the $d^{2}C_{v}/dT^{2}$ verses $T$ curves of the model
in \cite{7} (Pagnani et al) which discusses a disordered (glassy)
statistical model of RNA secondary structures. The specific heat
analysis visibly presents the peculiar differences in the
characteristics of the two models at small and large $N$'s with
$N=1$ showing an unusual double peaked behavior.

\begin{figure}
\includegraphics[width=6cm]{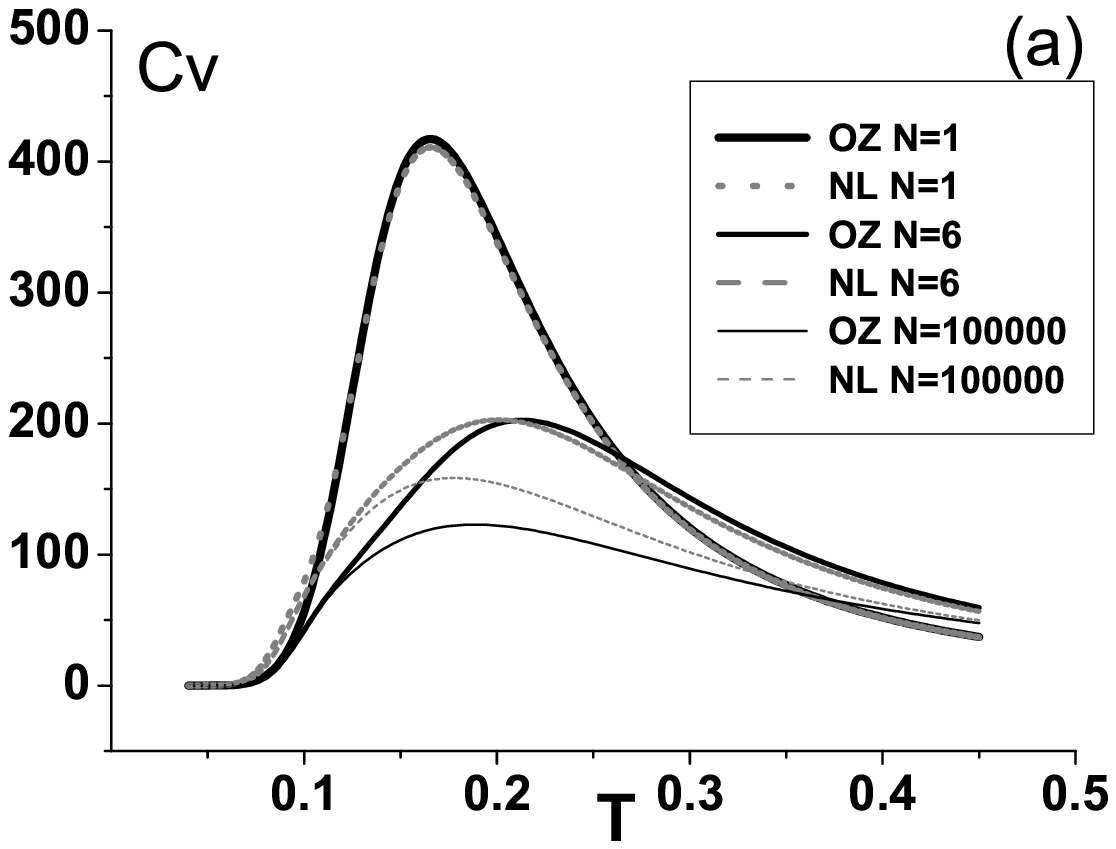}
\includegraphics[width=6cm]{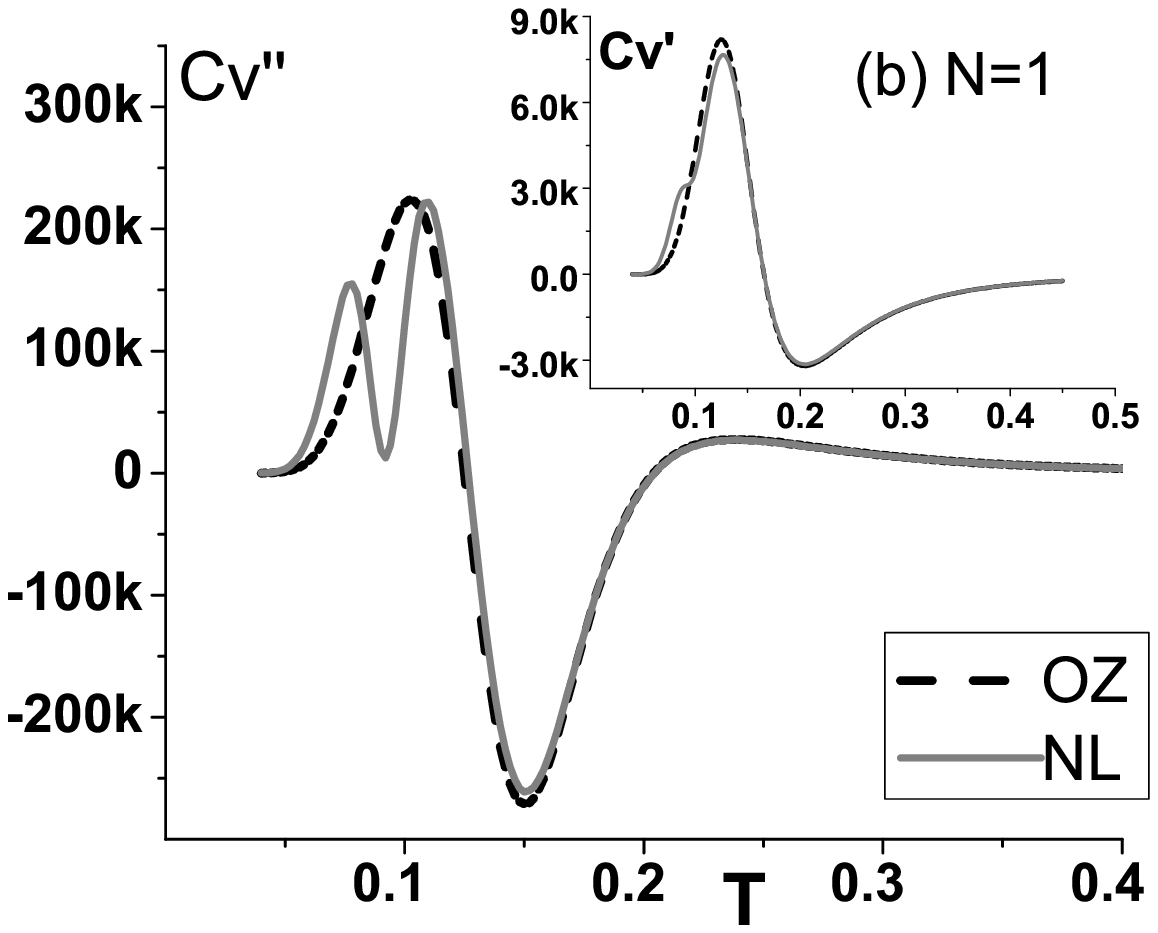}\\
\includegraphics[width=6cm]{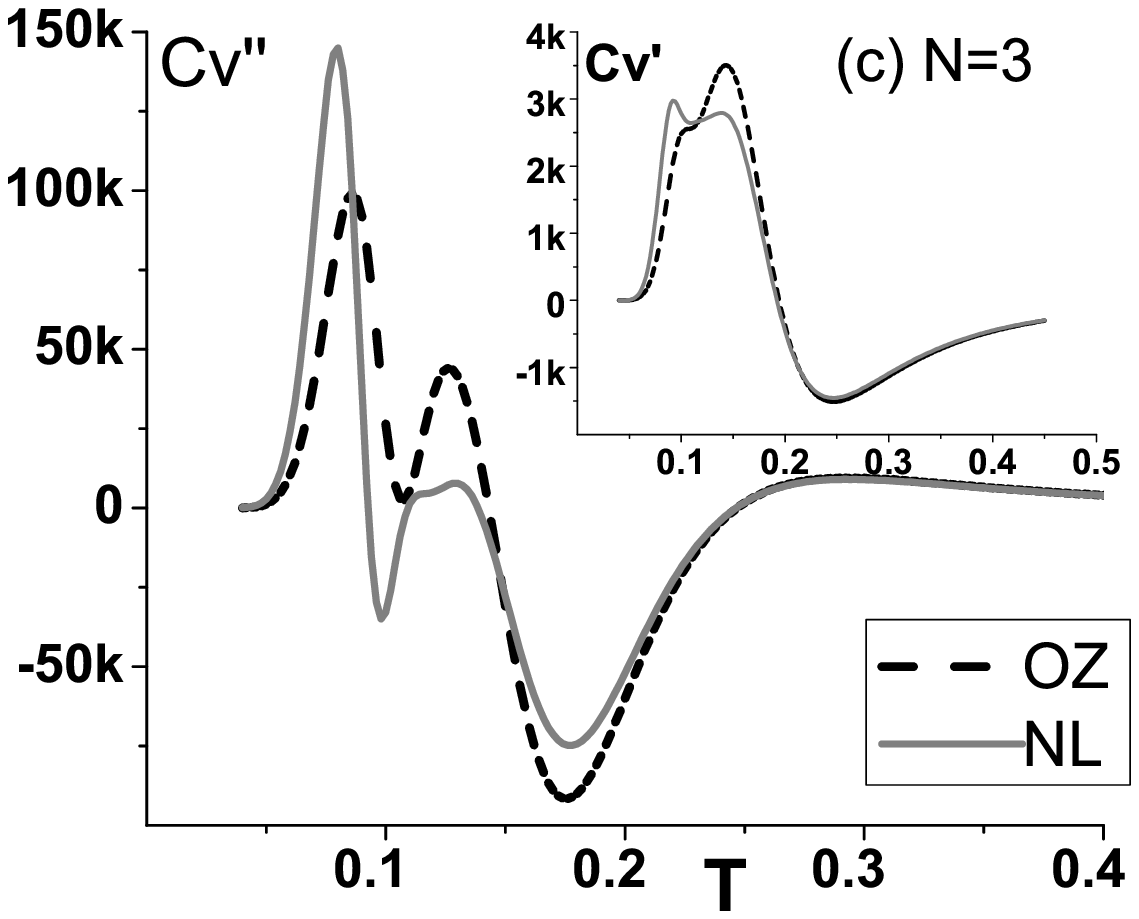}
\includegraphics[width=6cm]{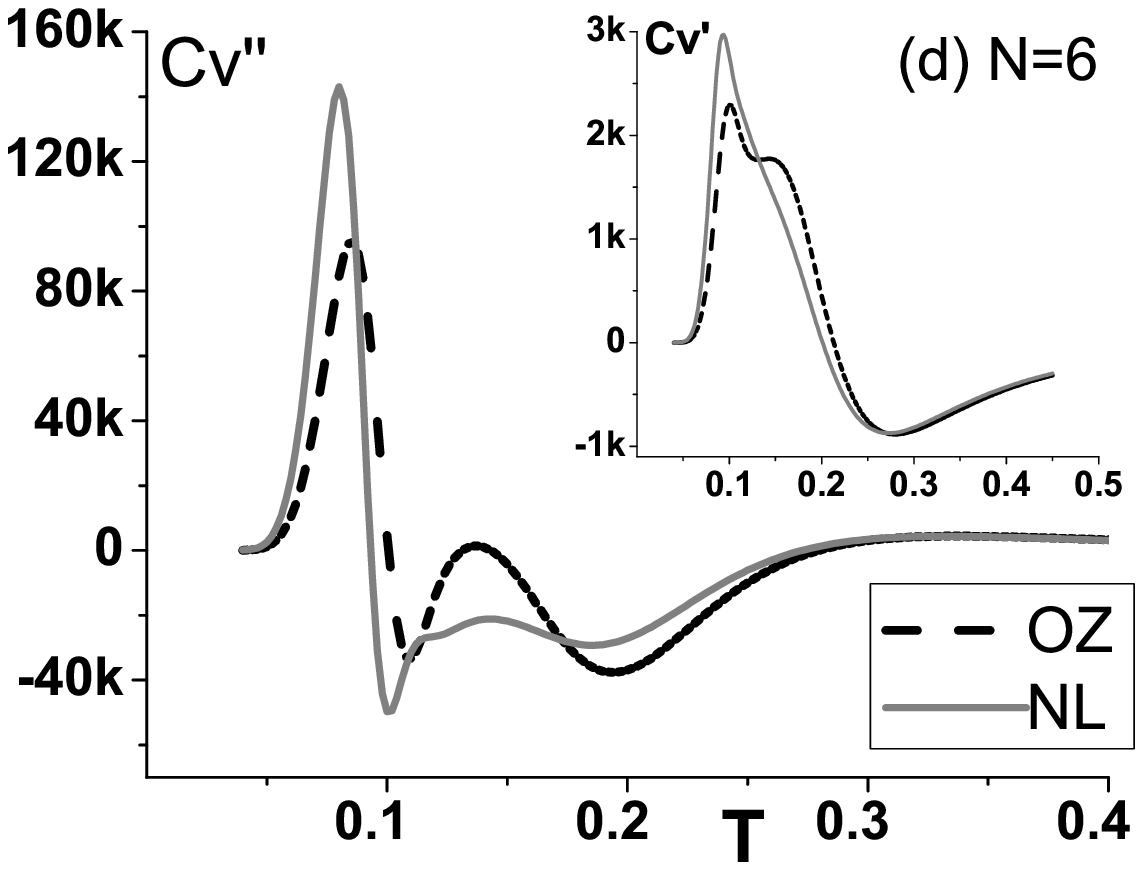}\\
\includegraphics[width=6cm]{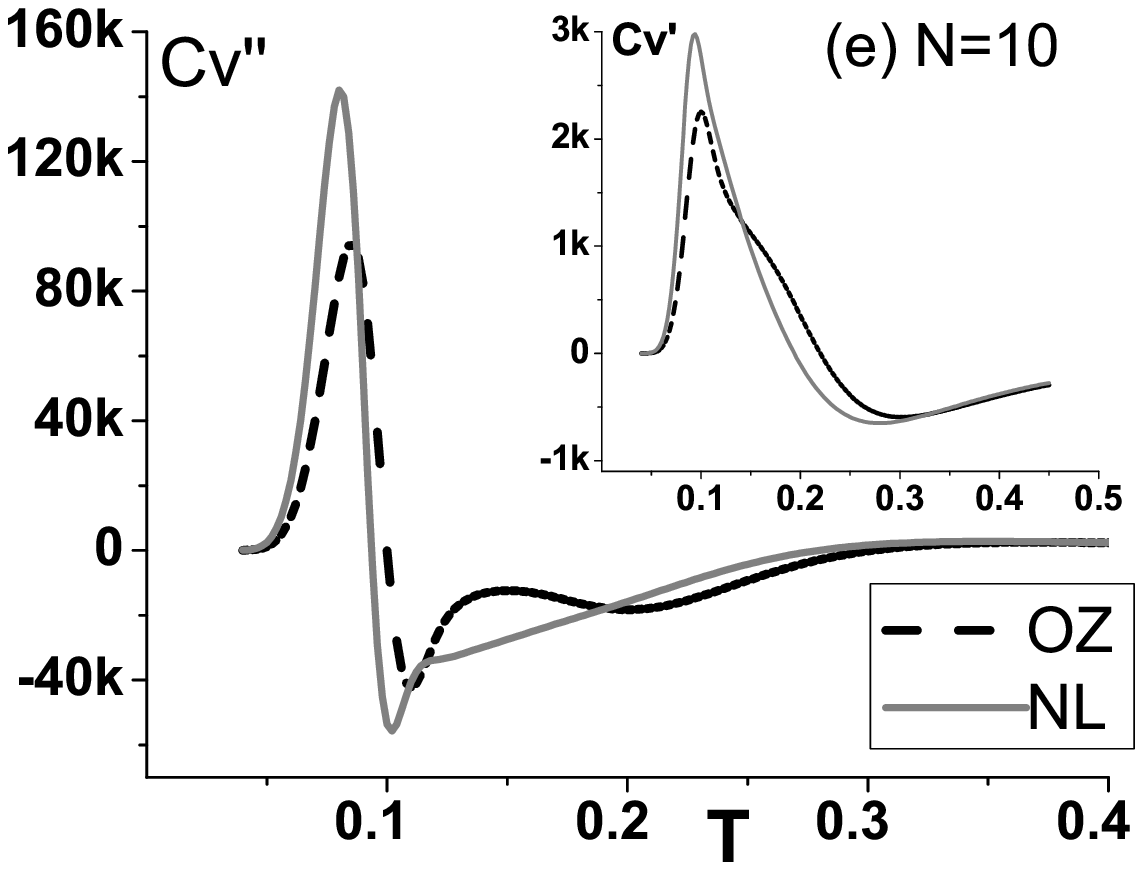}
\includegraphics[width=6cm]{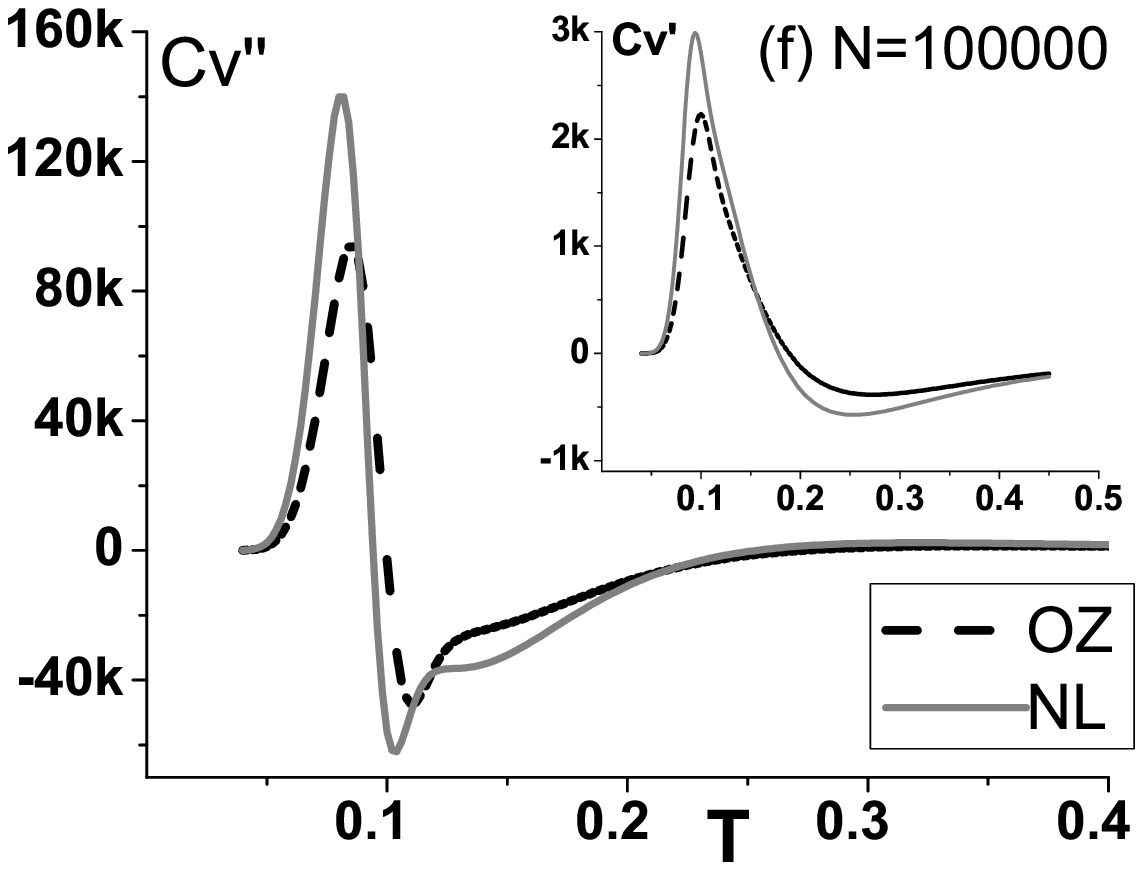}
\caption{The figure shows specific heat $C_{v}$, $C^\prime_{v}$
(inset) and $C^{\prime\prime}_{v}$ as a function of $T$ for the
logarithmic interaction (solid line) and $oz$ models (dashed line)
for different values of $N$.}
\end{figure}

\section{Tapping}
In tapping, the matrix is coupled to an external
source, which is then removed and the number of different
configurations (multiple solutions) are counted (Deo \cite{9}). The
limit of external source $\rightarrow 0$ gives different moments
which may result in different partition functions and hence free
energies (considering different tappings explores entire space of
the configurations). For the log model, the form of potential is
$V(\sigma)=[\frac{\mu}{2}\sigma^{2}-\frac{t^{\prime\prime}}{2} (\log
\sigma^{2})]$ with $\lambda_{i}$ as the eigenvalues of $\sigma$.
Introducing a linear matrix source $\sigma/y$ in the action of Eq.
(1) (after solving), the saddle points are found from the equation
$\frac{\partial [V(\lambda_{i})-\frac{1}{y}\lambda_{i}]}{\partial
\lambda_{i}}=0$ to be 

\begin{equation}
\sigma_{c}=\frac{\frac{1}{\mu y}\pm
\sqrt{\frac{1}{\mu^{2}y^{2}}+\frac{4t^{\prime\prime}}{\mu}}}{2}
\end{equation}

where $\mu=1/v$ and $t^{\prime\prime}=1/w$. Depending upon small or
large $1/y$, the saddle points are 

\begin{eqnarray}
\sigma & = & + \sqrt{\frac{t^{\prime\prime}}{\mu}}+ \frac{1}{2}(\frac{1}{\mu w}) \; ; \; \frac{1}{y} > 0. \\
\sigma & = & - \sqrt{\frac{t^{\prime\prime}}{\mu}}+ \frac{1}{2}(\frac{1}{\mu w}) \; ; \; \frac{1}{y} < 0. 
\end{eqnarray}

The action $S(\sigma)=\frac{\mu}{2}\sigma^{2}-\frac{t^{\prime\prime}}{2}log(\sigma^{2})-\frac{1}{y}\sigma$
therefore becomes

\begin{equation}
S(\sigma^{\pm}_{c})=\frac{t^{\prime\prime}}{2}[1-log(\frac{t^{\prime\prime}}{\mu})]\mp
\sqrt{\frac{t^{\prime\prime}}{\mu}}\frac{1}{y}. 
\end{equation}

Thus the set of moments grows exponentially as $2^{N/2}$. This suggests the
intriguing possibility that log interaction matrix model (and also $oz$)
represent a class of `glassy' matrix models.

\section{Conclusions}
The Letter studies a Penner type logarithmic
interaction in the framework of a random matrix model of RNA folding
and its structure combinatorics. This is the first instance where
Penner matrix models have been applied to the study of RNAs. The
asymptotic analysis suggests: (i) There exists different regions of
$N$ in which the structural properties of the log interaction model
(and also the $oz$) are different. (ii) A dependence of the combinatoric factor on the matrix size $N$
is found for the matrix models of RNA with and without interactions
which has not been found before. (iii) The effect of interaction
is visible in $v$ and $N$ which is
given by the re-scaling of $v$ to twice that in the $oz$ (for large
$N$ region) and dominated by $N$ (in the small $N$ region). The
substitution $v/2$ in the log partition function at large $N$,
reduces the structures to that found for the $oz$ model. (iv) The
$L^{3g-3/2}$ behavior for large $N$ in the log interaction model is
in good agreement with the $oz$ model. For small $N$, these exponents
change as the secondary region is stretched longer while the
tertiary region gets shortened. The effect of
interaction is thus visible in the way the structures are
distributed for a given length which may represent a new universality class, Fig. 1(b) \cite{8}.
The specific heat clearly
highlights the different structural behavior in the two regions of
$N$ with an unusual double peak for $N=1$ and a kink for large $N$.
The tapping explicitly shows the presence of multiple solutions in
these matrix models. An important necessary direction in these
models is the derivation of the generating function with any
$N^\prime t^\prime$. This will help clarify some unresolved issues
such as the genus characterization (which at the moment is
indistinguishable from effect of interaction both of which are
contained in $N$). The matrix model of RNA with logarithmic interaction provides a
larger class of interacting RNA structures than
the matrix model of \cite{3,4,5,6}. This is because an additional (and
therefore complete) dependence of the structures on $n$ (other than vertices and edges) in
the model will be possible in terms of all the $n$ punctures of the
Riemann surfaces.

We thank Professor Henri Orland. We thank CSIR and UGC for senior and junior research fellowships
and the University faculty R\&D research program for financial support.

\end{document}